# Charged-particle control via spatio-temporally tailored pulses from gas-based nonlinear optics

HAO ZHANG,[1,2] JOSHUA MANN,[3] JAMES ROSENZWEIG,[3] MICHAEL CHINI,[4] AND SERGIO CARBAJO[1,2,3,5,*]

[1]Department of Electrical & Computer Engineering, University of California, Los Angeles, Los Angeles, CA 90095, USA
[2]SLAC National Accelerator Laboratory, Stanford University, Menlo Park, California 94025, USA
[3]Physics and Astronomy Department, University of California, Los Angeles, CA 90095, USA
[4]Department of Physics, The Ohio State University, Columbus, Ohio 43210, USA
[5]California NanoSystems Institute, Los Angeles, CA 90095, USA
haozh@g.ucla.edu; scarbajo@g.ucla.edu

**Abstract:** Gas-filled waveguides enable few-cycle, spatio-temporally coupled (STC) pulses with programmable structure, opening new routes to control charged particles with optical fields. This review maps the landscape of optical-field-driven photoemission, then surveys gas-based nonlinear drivers, photonic crystal fibers (PCFs) for low-energy, high-repetition operation and hollow-core capillaries (HCCs) for high-power, few-cycle synthesis. We highlight mechanisms for deterministic pulse shaping, including four-wave-mixing-based spectral-phase transfer in HCCs, and show how tailored STC waveforms steer emission dynamics from the multiphoton to tunneling regimes, enabling sub-cycle gating, momentum control, and brightness scaling. We conclude with open challenges: phase stability, mid-IR scalability, coupling to nanophotonic emitters, metrology of vectorial fields, and outline a path toward compact, ultrafast, phase-coherent electron sources and emerging quantum applications powered by nonlinear photonics.

## 1. Introduction

Controlling charged particles lies at the heart of modern ultrafast science and technology, spanning ultrafast electron diffraction [1,2], compact accelerators [3,4], and attosecond physics [5,6]. Traditionally, this has relied on static or radiofrequency (RF) electromagnetic fields, which have enabled tremendous progress in charged-particle optics and accelerator physics [7,8]. However, those approaches are intrinsically constrained by slow temporal response, poor spatial resolution, and the challenge of scaling down to tabletop ultrafast platforms [9,10]. Herein, there has been a growing demand for novel methods that can offer compact, efficient, and programmable ultrafast control of electrons and ions [11,12].

In recent years, spatio-temporally controlled (STC) laser pulses have emerged as a transformative tool for charged-particle control [13–15]. By manipulating the amplitude, phase, polarization, and spatio-temporal coupling of picosecond-to-attosecond pulses, it becomes possible to shape photoemitted wave packets and exert forces on free electrons with high precision. Such optical-field-driven control could enable attosecond streaking of electrons [16–18], ultrafast switching of photocathodes [19,20], direct-field laser acceleration in vacuum [21–23], and coherent acceleration in dielectric or nanoplasmonic structures [24,25]. STC pulses also allow simultaneous control over both the temporal and momentum distributions of photoemitted electrons, resulting in coherent electron sources for next-generation imaging and quantum technologies [26–30].

The implementation of STC pulses for charged-particle control highly depends on ultrafast nonlinear optics [14,31–33]. Among various platforms, gas-based nonlinear optics in waveguides has proven uniquely powerful, combining broadband spectral broadening with high transmission and tunable dispersion engineering [34–36]. Gas-filled photonic crystal fibers (PCFs) and gas-filled hollow-core capillaries (HCCs) have become popular platforms for generating few-cycle and even sub-cycle pulses across the UV to mid-infrared regimes [35,37–41].

PCFs filled with noble gases have enabled the generation of broad supercontinuum spectra spanning from the visible to the mid-infrared, with exceptional spectral coherence and control over phase-matching conditions [35,42]. These platforms support the development of soliton-driven dynamics [37,38,43], dispersive wave emission (RDW) [43–46], and tunable spectral shaping by exploiting the interplay between self-phase modulation, molecular Raman response, plasma effects, and waveguide dispersion. Recent works have also utilized PCFs to compress pulses to sub-10 fs durations at nanojoule-to-microjoule energy levels, making them ideal candidates for driving attosecond streaking, ultrafast photoemission studies, and low-energy strong-field experiments in compact settings [47,48].

In parallel, HCC systems filled with controlled gas mixtures have achieved multi-millijoule, carrier-envelope phase (CEP) stable, and few-cycle pulse outputs [38,41,49–51]. Such platforms are now widely employed to drive high-harmonic generation (HHG) in gases and solids, enabling both attosecond pulse trains and isolated attosecond pulses across a wide photon energy range, from the extreme ultraviolet (XUV) to soft X-rays [52,53]. These systems also serve as backend compressors for high-power OPCPA, Yb: solid-state, or Ti: sapphire lasers, forming the basis of many state-of-the-art attosecond beamlines [54–56].

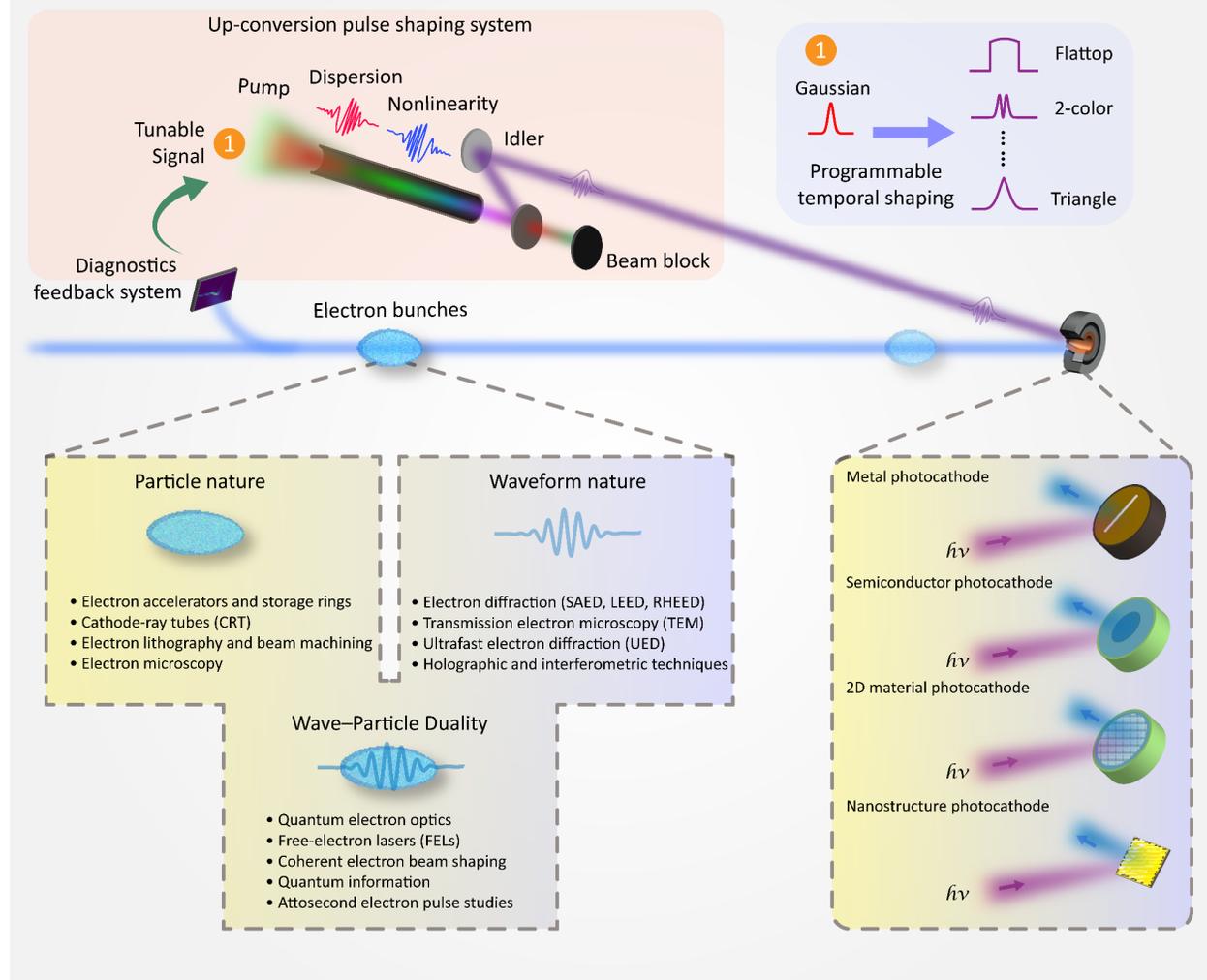

Figure 1. Schematic overview of parametric-frequency-conversion-based temporal shaping for photoemission and its implications for electron beam control. A tunable signal undergoes nonlinear up-conversion with a shaped pump pulse to generate a broadband idler pulse with user-defined temporal profiles (e.g., Gaussian, flattop, 2-color). The shaped ultraviolet light illuminates various types of photocathodes—metal, semiconductor, 2D material, and

nanostructured—producing customized electron bunches. The emitted electrons can be tailored for specific applications depending on whether their particle nature, wave nature, or wave–particle duality is exploited. Applications span particle acceleration, ultrafast diffraction and microscopy, and emerging quantum electron optics.

Despite those achievements, the interplay between gas dynamics, nonlinear propagation, and spatio-temporal couplings is still not fully understood, especially under high-repetition-rate[57] and high-energy driving conditions [27]. Integrating these optical-field control with advanced photoemitter materials, such as 2D materials [58,59], nanostructures [60,61], and quantum emitters [62,63], requires novel strategies to ensure stability, scalability, and high reproducibility. Figure 1 provides an overview of the integrated landscape for shaping electron beams using up-conversion-based pulse shaping and tailored photocathodes. It highlights how programmable spectral-temporal shaping, via dispersion and nonlinearity in optical up-conversion, can be combined with a range of photocathode platforms to generate electron pulses with tunable structure and coherence. The emitted electron wavepackets support applications across a broad spectrum of regimes, from classical particle acceleration and microscopy to coherent electron beam manipulation and quantum information protocols.

In this review, we summarize recent advances in charged-particle control enabled by STC pulses generated and driven from gas-based nonlinear optical systems. We first provide an overview of photoemitter platforms relevant to optical-field-driven emission. We then examine gas-filled waveguides as drivers of ultrafast nonlinear optics, between the low-energy PCF approaches and the high-power HCCs techniques. Finally, we discuss open challenges and prospects, outlining the path toward compact, ultrafast, and coherent electron sources powered by nonlinear photonics.

## 2. Photoemitters overview

This section surveys the emitter landscape, clarifies operating regimes and figures of merit, and highlights how STC pulses couple to photoemission dynamics.

### 2.1 Classes of Photoemitters

A wide range of materials has been explored as photoemitters, each suited to distinct applications. Metals such as gold, silver, and tungsten remain robust and stable under intense fields [10,64]. Their relatively high work functions (~4 eV), however, often necessitate ultraviolet or high-intensity (i.e., multi-photon) near-infrared excitation.

Semiconductors offer lower work functions and tunable band structures, enabling more efficient emission at moderate optical intensities. They have become central to ultrafast electron diffraction and photocathode-based accelerator technologies [65–67]. More recently, low-dimensional systems, including graphene [68,69] and transition-metal dichalcogenides [70,71], have attracted strong interest. Their atomic thickness and reduced effective work functions provide enhanced nonlinear optical response, allowing tunneling emission with modest few-cycle infrared drivers [72,73]. Nanostructured emitters such as sharp metallic tips [74,75] or nanoplasmonic arrays [75] push local field strengths to extreme levels, reducing emission thresholds by orders of magnitude and enabling attosecond-scale electron bursts with directional coherence [74].

To place these classes in context, **Table 1** summarizes representative emitter materials, including quantum efficiency (QE), operating wavelength, vacuum requirements, and acceleration gradients.

Table 1 Representative photocathode materials and their photoemission characteristics

| Photocathode | Type | QE (%) | λ_op (nm) | Work Function / Threshold |
|---|---|---|---|---|
| Cu [76] | Metal | ~1×10⁻⁵ | 253–266 | 4.6–4.9 eV |
| Mg [77] | Metal | 1×10⁻³–3×10⁻³ | 262 | 3.7 eV |
| Pb (bulk/film) [78] | Metal | 5×10⁻⁴–3×10⁻³ | 193–213 | 4.1 eV |
| Nb [79] | Metal | 5.7×10⁻⁶ | 258 | 4.3 eV |
| Cs$_2$Te [80] | Semiconductor | 3–10 | 257–262 | 3.5 eV |
| CsK$_2$Sb [81] | Semiconductor | 4–10 | 520 | 1.6–2.0 eV |
| Cs$_3$Sb / NaKSb [82] | Semiconductor | 2–5 | 515–520 | 1.6–2.1 eV |
| Graphene [69] | 2D material | ~0.1 | 800–2000 | 3.0-4.0 eV |
| MoS$_2$, WS$_2$ [10] | 2D material | <1 | 400–800 | 4.0–5.0 eV |
| Carbon nanotubes [83] | Nanostructure | <0.1 | 400–1000 | 4.0–5.0 eV |
| Au nanotips [84] | Nanostructure | ~10⁻⁴ (field-driven) | 700–2000 | 4.7–5.4 eV |

## 2.2 Emission and operation regimes

The operating regime of photoemitters depends critically on the interplay between laser parameters and material response. A convenient framework is provided by the Keldysh parameter γ [85],

$$\gamma = \sqrt{\Phi/2U_p} = \frac{\omega\sqrt{2m\Phi}}{eE_0},$$

where Φ is the work function (or electron affinity), $U_p = e^2 E_0^2 / 4m\omega^2$ the ponderomotive potential (the average quiver energy of the electron in the optical field), ω the laser angular frequency, $E_0$ the peak electric field, and *m*, *e* the electron mass and charge. This dimensionless parameter compares the ponderomotive potential to the work function. It therefore distinguishes between multiphoton -dominated emission (γ≫1, small ponderomotive potential) and optical-field-driven tunneling emission (γ≲1, large ponderomotive potential). In all cases, multiple pathways produce current simultaneously, including through above threshold photoemission (ATP) where more than the necessary number of photons are absorbed, and photo-assisted field emission, where fewer than the necessary number of photons are absorbed [86]. These mechanisms are depicted in Fig. 2.

In the multiphoton regime, electrons overcome the potential barrier by absorbing $N = \lceil \Phi/\hbar\omega \rceil$ photons. Assuming the probability of absorbing each photon is constant, the emission rate follows the well-known power law:

$$W_{MPPE} \propto I^N,$$

with $I$ the laser intensity. This mechanism is stable and widely exploited in metallic emitters, but it typically produces relatively broad energy spectra due to the probabilistic nature of photon absorption.

As the optical intensity increases, so does the ponderomotive potential and therefore the energy of a free electron state. Specifically, the Volkov state's energy above the vacuum level is the ponderomotive potential. This ultimately increases the effective work function, with the number of photons now required to photoemit being $N = \lceil (\Phi + U_p)/\hbar\omega \rceil$, prohibiting fewer-photon channels from being utilized. This diminishes the yield to below the expected power law behavior and is known as channel closing [87].

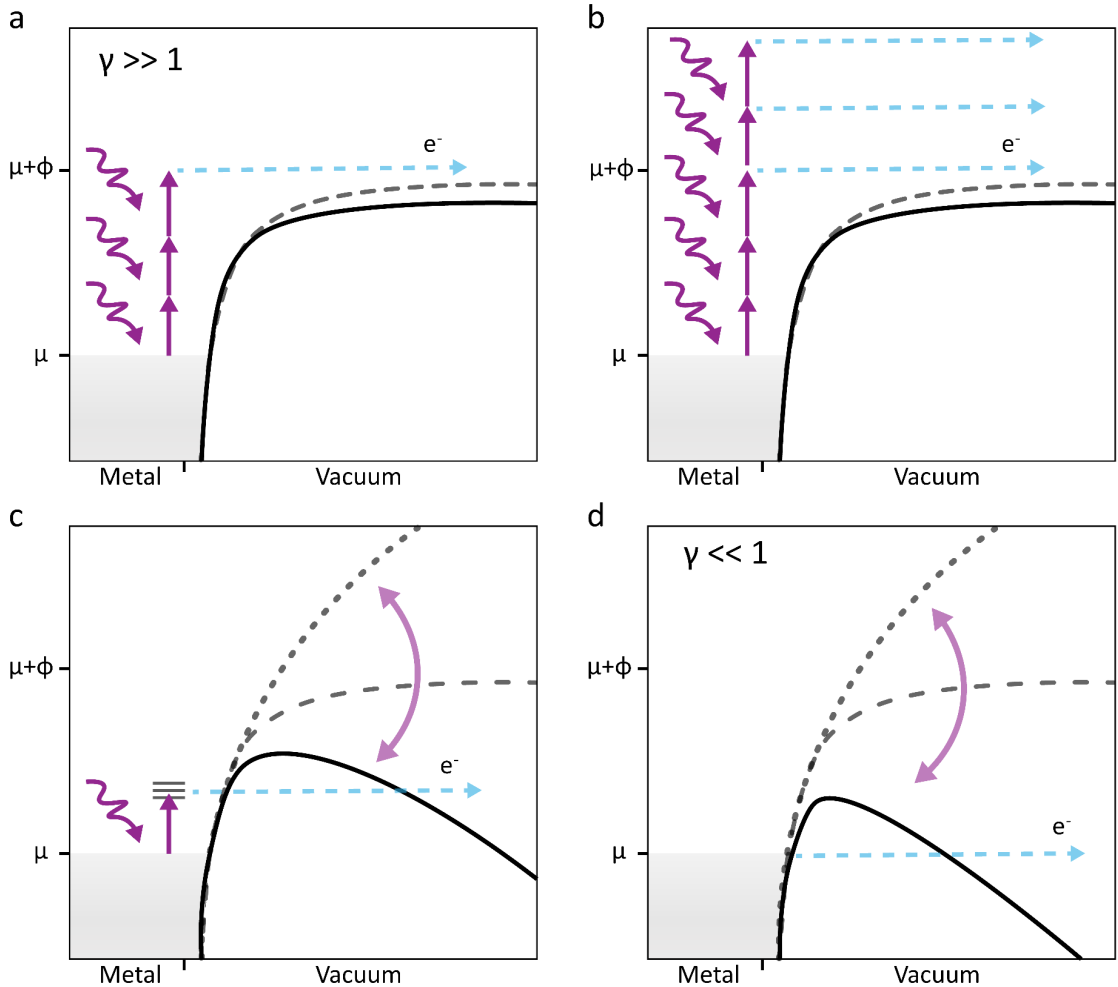

Figure 2. Potential-energy schematics of photoemission from a metal under the single active electron approximation and the dipole approximation [86], in order of increasing field sensitivity: (a) multiphoton photoemission; (b) above-

threshold photoemission; (c) photofield emission; (d) optical field emission. For low-field cases, the electrostatic potential (dashed gray) is negligibly lowered (solid black) or raised (fine-dashed gray) under the optical field and the dominant emission mechanism is photon absorption. For high-field cases, the potential is appreciably lowered, resulting in sub-cycle quantum tunneling and field emission.

With even further increasing optical intensities, the dominant picture becomes quasi-static field emission rather than photoemission under three conditions. The magnetic component of the optical field must be negligible, which is the case when the magnetic displacement, $\beta_0 = U_p/2mc\omega < 1\ a.u.$, and the radiation pressure, $T_{rad} = U_p^2/2mc^2 < 1/2\ a.u.$, are small. In such a case, a single-active-electron model may interact with the optical field using the length gauge, $V = eE_0 \cdot r$, placing us within the "dipole oasis" and permitting this dipole approximation. Then, within the tunneling regime, $\gamma \lesssim 1$, and with a sufficient frequency such that this dipole approximation may be applied, we are then within the "tunnel oasis" where the Fowler-Nordheim (FN) framework of (quasi-)static field emission may be utilized [88]. Driven by CEP-stable few-cycle pulses, the resulting tunneling emission gives rise to attosecond-resolved electron bursts, opening the door to coherent control of photoemitted wave packets.

Within the FN framework, the applied optical field suppresses the vacuum potential in half of the optical cycle, enabling sub-cycle electron emission via quantum tunneling, constituting field emission. The tunneling rate is often described by an FN-like expression [89,90]:

$$W_{tunnel} \propto exp\left[-\frac{4\sqrt{2m}}{3\hbar eE_0}\Phi^{3/2}\right],$$

which highlights the exponential sensitivity to both the applied field strength and the work function. While this particular result follows from a simplistic "exact triangular" surface potential model, there are other surface potential models (e.g., the Schottky-Nordheim barrier [91], which admits the Schottky effect), evaluation approaches (from the Wentzel–Kramers–Brillouin and small decay width approximations [92] to direct numerical evaluation [93]), and the inclusion of other phenomena such as the field enhancement factor which further increases the apparent surface field.

As the total liberated charge increases, the collective restoring force diminishes the yield due to the space charge of the emitted electrons and the image charge left on the cathode. Three factors assist the emission of charges by countering these forces: (1) the externally applied bias field or RF field, (2) non-zero emission energy, and (3) the ponderomotive force. Ignoring the externally applied fields, both the emission energy and the ponderomotive force scale linearly with the ponderomotive potential, which is linearly proportional to the optical intensity. Therefore, since the image force (or potential) scales with the total charge, the emitted charge is also proportional to the optical intensity. This reduces the power law from the low-field multiphoton regime, $n \propto I^N$, to this high-field space-charge (or image-charge) limited regime, $n \propto I$ [94].

These extreme regimes need not necessarily be achieved directly with a free optical field. Nanostructured and low-dimensional emitters are ubiquitous for effectively reducing the local work function and enhancing the optical field intensity. Plasmonic tip structures, for example, can provide local enhancement factors of 10–100, lowering the effective Keldysh parameter and allowing tunneling-like emission under moderate free laser intensities [3,74,75,84]. This plasmon-assisted regime has been exploited for attosecond streaking and sub-cycle control of electron emission.

### 3. Gas-filled waveguides as nonlinear optical drivers

Gas-filled waveguides provide a unique and tunable platform for generating STC pulses, enabling control over spectral bandwidth, temporal duration and spatial mode quality [35,37]. Compared with bulk nonlinear crystals, gas-based systems offer significant advantages, including broader transparency windows, higher damage thresholds, and flexible dispersion engineering. These benefits are particularly pronounced in the deep ultraviolet (DUV) and vacuum ultraviolet (VUV) regimes, where crystal-based systems often suffer from severe optical damage, strong absorption, and limited phase-matching options. In contrast, gas-filled waveguides can sustain high intensities while maintaining phase coherence and low loss, making them indispensable for high-fidelity pulse shaping in these extreme spectral domains.

**3.1 Gas-filled PCF: low-energy ultrafast drivers**

Gas-filled anti-resonant guiding photonic crystal fibers (AR-PCFs) represent a highly effective platform for ultrafast nonlinear optics. Unlike photonic bandgap fibers, which suffer from narrow transmission bandwidths. AR-PCFs support ultrabroadband, low-loss transmission[95], making them well-suited for few-cycle pulse manipulation. Their microstructured claddings provide strong modal confinement in a central hollow core filled with noble gases such as Argon, Neon, or Helium. This geometry reduces the nonlinear threshold by orders of magnitude compared to free-space gas cells, enabling nonlinear interactions at input energies from tens of nanojoules to a few microjoules. For example, kagomé-lattice PCFs filled with Argon have compressed 50-fs Ti: Sapphire pulses down to 2.4 fs, while preserving good CEP stability [96]. Such performance would be unattainable in bulk or free-space gas systems at similar energy levels.

The physics of gas-filled PCFs is governed by the interplay of waveguide dispersion, Kerr nonlinearity, and plasma effects. When operated in the anomalous-dispersion regime, soliton dynamics dominate: high-order solitons undergo self-compression to durations shorter than one optical cycle, often reaching sub-femtosecond scales when driven at 800 nm, before undergoing fission into multiple fundamental solitons and dispersive waves. This mechanism underlies supercontinuum generation in the visible to mid-infrared. For instance, Ermolov *et al.* demonstrated a three-octave SC spanning 112 nm to 1000 nm in helium-filled kagomé PCFs using sub-μJ input energies [97]. When the same class of fibers is filled with argon at elevated pressures, phase-matched RDW emission occurs in the deep UV up to few MHz repetition rates [35,98]; tunable narrowband pulses between 200 and 600 nm driven by a few-μJ pump have been demonstrated [99]. These capabilities make PCFs uniquely powerful for simultaneously accessing the UV, visible, and IR with high coherence. PCFs also enable mid-infrared ultrafast sources that are particularly relevant for strong-field photoemission [100]. These properties translate directly to applications in charged-particle control. Few-cycle PCF outputs have been employed to drive strong-field photoemission from metallic nanotips, where the timing and angular distribution of emitted electrons can be influenced by the pulse waveform. Similarly, broadband SC and dispersive wave outputs enable two-color schemes: combining, for instance, an infrared driver with a UV control field to manipulate photoelectron spectra and emission timing in semiconductors and 2D materials. Because the required input energies are modest, such platforms operate efficiently at high repetition rates, enabling compact tabletop experiments that bridge nonlinear photonics with ultrafast electron science.

**3.2 Gas-Filled HCF: High-Energy Few-Cycle Drivers**

Unlike PCFs, which use microstructured claddings to enable strong modal confinement and flexible dispersion engineering, HCCs are simple dielectric waveguides, typically 50–500 μm in diameter and up to several meters long, with smooth inner walls and large mode areas. While both PCFs and HCCs can support soliton dynamics under appropriate conditions, HCCs are more commonly used for spectral broadening via self-phase modulation (SPM) and post-compression due to their simplicity and high transmission. When filled with noble gases at controlled pressures (up to several bars), they support efficient spectral broadening via SPM while maintaining low loss and high damage tolerance. The large core area reduces nonlinear accumulation and allows millijoule to hundred-millijoule-level pulses to be transmitted and compressed without optical damage [38,45].

The dominant mechanism for pulse compression in gas-filled HCCs is SPM–driven spectral broadening, followed by dispersion compensation. Early studies already showed that Ti: Sapphire pulses with millijoule-level energies could be compressed below 12 fs in noble gas-filled capillaries [101,102]. Recent advances have pushed the performance envelope further. For instance, sub-cycle field transients and optical attosecond pulses have been synthesized from multi-octave-spanning supercontinua generated in gas-filled waveguides [103,104]. Soliton-assisted self-compression has enabled the efficient generation of few-cycle pulses with microjoule to millijoule energies, often accompanied by simultaneous dispersive-wave emission into the ultraviolet. In recent implementations, this approach has demonstrated the generation of sub-1 fs pulses at multi-millijoule energies, as well as DUV dispersive-wave pulses exceeding 100 μJ, opening new frontiers in ultrafast spectroscopy and attosecond science[105].

A unique strength of HCCs is their wavelength scalability. By varying the gas species and pressure, broadening and compression can be optimized across both the near-IR and mid-IR ranges [106–110]. The use of molecular gases [111–113] or shaped input pulses [114] can further enable overall shifts of the central frequency. These capabilities are crucial because the HHG cutoff [115] and the efficiency of sub-cycle photoemission both benefit from longer driving wavelengths [52]. More recently, soliton-based approaches to HCC-based compression have enabled the measurement of isolated attosecond pulses with 350 as duration within the DUV region [116]. Such developments significantly expand the functionality of hollow-core systems as programmable platforms for ultrafast electron dynamics and light–matter interaction studies, with capabilities such as polarization control (including circular polarization and radial vector beams) offering new degrees of freedom for tailoring optical fields..

For charged-particle control, HCCs extend the capabilities of gas-based nonlinear optics into the high-energy regime where dense electron bunches and relativistic dynamics become accessible. When multi-millijoule pulses are compressed to a few optical cycles, they can trigger strong-field photoemission from metallic and semiconductor surfaces, generating femtosecond electron bunches with charge densities far exceeding those achievable with PCF-based drivers. Such bunches are directly relevant for UED and time-resolved photoemission microscopy, where both temporal resolution and beam brightness are paramount.

### 3.3 Integration with spatio-temporal control

Beyond serving as broadband compressors, gas-filled waveguides also provide a natural platform for implementing STC of ultrashort pulses. By engineering effects such as wavefront rotation, pulse-front tilt, or spatial chirp, one can map femtosecond temporal structures into angular or spatial coordinates, enabling new schemes of photoelectron steering [117], attosecond gating [118], and atomic transition [119]. In dielectric nanostructures or plasmonic channels, the sub-cycle optical field can impart controlled momentum kicks to free electrons, effectively mapping optical phase into electron energy and angle [120].

A particularly promising direction is the use of mid-IR HCC drivers to explore relativistic electron control [121]. Because the ponderomotive potential scales as $U_p \propto I \lambda^2$, driving with 2–4 μm few-cycle pulses allows relativistic quiver motion at significantly reduced intensities [122]. This has enabled early demonstrations of lightwave acceleration over sub-millimeter distances, pointing toward compact optical accelerators where the temporal structure of the laser pulse is imprinted directly onto the electron beam. Such capabilities hint at the integration of HCC-based pulse compression with emerging platforms in dielectric laser acceleration and laser–plasma injectors, where precise control of electron emission and injection phases is essential.

Recent work has demonstrated a feasible way of controlling the spatio-temporal pulse shape by managing the spectral phase transfer between two wavelengths [15,27,123]. We investigated four-wave mixing (FWM) in gas-filled hollow-core capillaries as a mechanism for high-fidelity phase transfer from the signal to the idler field, as shown in Figure 3. This process offers a route to mapping spectral phase profiles into different frequency regions that are otherwise difficult to manipulate, such as the UV. The main challenge lies in the competition between linear dispersion and nonlinear dynamics, which tends to reduce transferred phase fidelity when conversion efficiency is high, and conversely suppresses conversion efficiency when fidelity is prioritized [27].

By exploring an intermediate energy regime, our study identified operational windows that balance these competing requirements. We showed that careful adjustment of the energy ratio between pump and signal, along with tailored pre-compensation of the input spectral phase, yields conversion efficiencies of 5–15% while preserving quasi-linear phase transfer. This allows, for example, femtosecond pulses shaped in the infrared to be upconverted into the UV with their spectral phase essentially intact. The ability to carry STC features across spectral bands through FWM thus establishes HCCs as more than compressors—they can function as waveform translators capable of generating multi-color, phase-coherent fields.

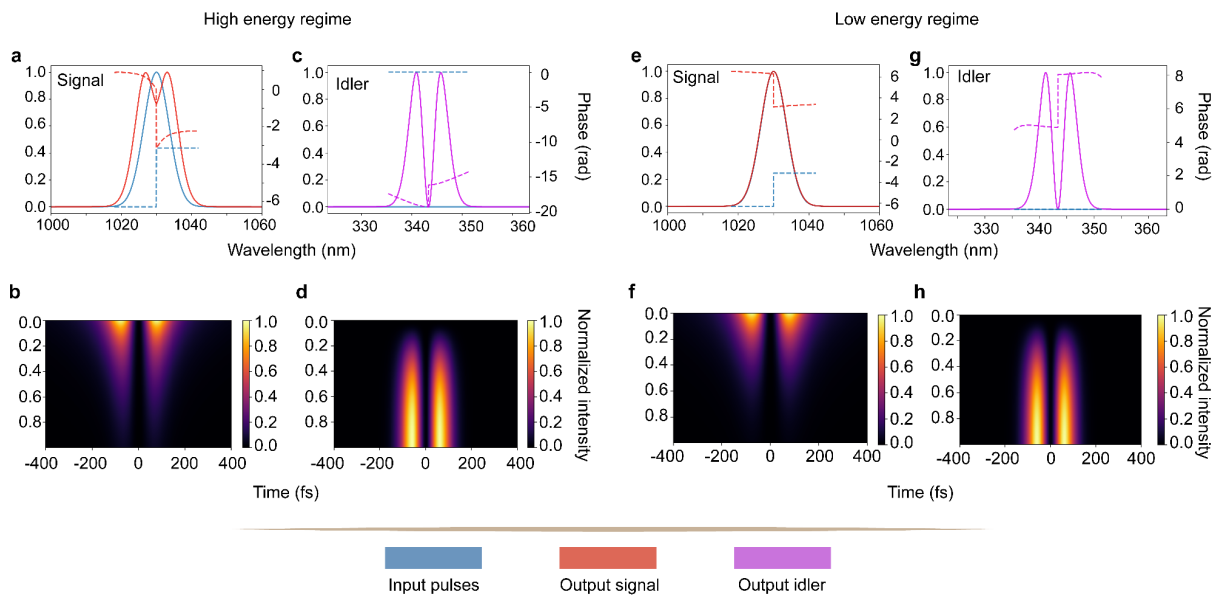

Figure 3. Simulation of spectral phase transfer in gas-filled hollow-core capillaries. High-energy regime (a–d): (a) input and output signal spectra and phase; (b) time-domain evolution of the signal pulse; (c) output idler spectra and phase; (d) time-domain evolution of the idler pulse. Strong nonlinear interactions distort the transferred phase and reduce conversion efficiency. Low-energy regime (e–h): (e) input and output signal spectra and phase; (f) time-domain evolution of the signal pulse; (g) output idler spectra and phase; (h) time-domain evolution of the idler pulse. At lower energies, nonlinear distortions are suppressed, leading to higher phase-transfer fidelity despite narrower idler bandwidths.

Spatio-temporal control can also be realized by taking advantage of multi-mode propagation in HCCs. For molecular gas-filled fibers, multimode propagation has enabled the generation of multi-dimensional solitary states (MDSSs) with few-cycle pulse durations and red-shifted frequencies [111]. Initially demonstrated in the infrared, the technique has since been used to produce ultrashort visible pulses with soliton-like profiles in space and time [124]. Multimode coupling also occurs in HCCs when there is substantial ionization. Recently, this effect has been leveraged to produce white-light supercontinuum pulses from 175 fs, 1 mJ laser pulses from a Yb:solid-state laser. Due to the interplay of nonlinearity, plasma, and multimode propagation, the pulses are compressed to <5 fs duration and >10 microjoule pulse energy directly at the fiber output [125].

The implications for charged-particle control are significant. UV–IR waveform pairs generated in this way enable two-color excitation schemes where the temporal gating or momentum shaping of electron emission is directly linked to engineered optical phases. Beyond electron science, the same principle is relevant to quantum transduction [126,127], precision spectroscopy [124], and attosecond metrology [118,128], where phase-coherent links between disparate frequency bands or entangled arrays are essential.

## 4. Discussion

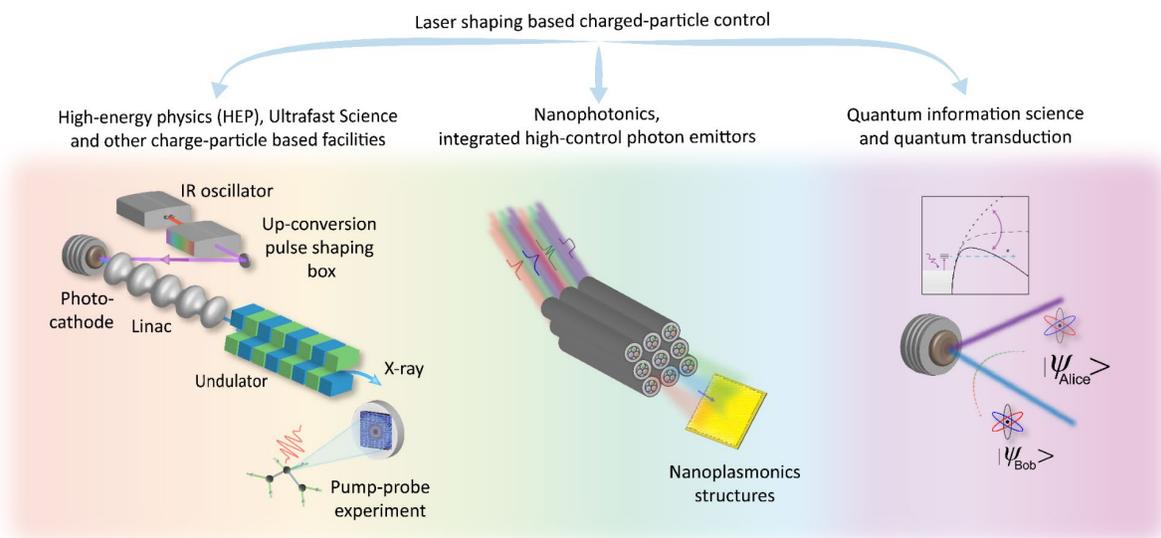

Figure 4. Applications of laser shaping-based charged-particle control across scientific disciplines. (a) In high-energy physics (HEP), ultrafast science, and accelerator-based light sources, shaped laser pulses from an IR oscillator are routed through an up-conversion pulse shaping box to control electron beams emitted from a photocathode. These beams are accelerated via a linear accelerator (Linac) and manipulated in an undulator to produce X-rays for pump-probe experiments in atomic and molecular systems; (b) In nanophotonics, arrays of structured light (e.g., carrying orbital angular momentum) are delivered through multichannel fibers to interact with engineered nanoplasmonic structures, enabling high-precision, on-demand single-photon emission; (c) In quantum information science and quantum transduction, photon-mediated interactions drive transitions in atomic systems, enabling entangled state preparation. The inset schematic of a Hamiltonian energy-level diagram demonstrates quantum-level transitions.

Programmable spatio-temporal shaping is emerging as a powerful tool for charged-particle control, bridging the gap between ultrafast optics and electron dynamics. The integration of these shaping techniques with gas-based nonlinear optics offers a uniquely versatile platform that spans energy scales from nanojoules to multi-millijoules, and time scales from femtoseconds to attoseconds. This capability enables not only the generation of tailored optical waveforms but also their direct imprinting onto free-electron wavepackets through mechanisms such as strong-field photoemission, optical streaking, and waveform-driven acceleration. In particular, gas-filled waveguides provide a programmable interface for both temporal and spatial control. Their broadband, high-damage-threshold nature allows for compression and shaping of pulses across diverse spectral regions, including the visible, near-IR, mid-IR, and even UV via nonlinear conversion. When combined with spatial structuring, these waveforms become capable of mapping temporal phase into transverse momentum or spatial emission profiles.

Looking ahead, several exciting directions emerge. One major avenue is the advancement of adaptive and feedback-optimized laser shaping techniques. Leveraging machine learning, deep learning, and real-time diagnostics, such as digital twin models illustrated in Figure 4a, will enable highly customized lightwave control over complex photoemission systems. These systems may include low-dimensional materials and strongly correlated electron systems, where conventional shaping approaches often fall short [11,129]. Another important direction involves multi-channel shaping architectures, as shown in Figure 4b. Platforms using PCFs or HCCs, potentially with independent phase modulation across channels, can support spatially multiplexed control. This opens the door to parallel electron beam modulation and multi-source injection schemes, significantly enhancing scalability and flexibility for beam-based applications [130]. Extending gas-based spatiotemporal control (STC) into the realm of quantum optics, as depicted in Figure 4c, presents exciting new frontiers. Shaped pulses in this regime can be used for coherent manipulation of light–matter interactions at the single-photon and few-photon level [18,131,132]. By precisely tailoring spectral phase across bands, one can envision novel routes for quantum transduction, entangled photon generation, and phase-coherent interfaces between photonic and electronic qubits. These capabilities establish gas-filled waveguides as powerful and versatile platforms in the broader landscape of quantum science [133]. The fusion of programmable optics, waveguide-enhanced nonlinear interactions, and charged-particle physics will enable a new class of coherent light–matter interfaces, where electrons and photons are synchronized on sub-femtosecond time scales and nanometer spatial scales. This synergy holds promise for transformative applications in ultrafast microscopy, quantum transduction, compact accelerators, and time-resolved electron imaging of matter in motion.

## 5. Conclusion

In this review paper, we have highlighted how gas-based nonlinear optics in hollow waveguides provides a powerful platform for integrating STC with charged-particle science. We discussed the operation regimes of photoemitters, the use of PCFs and HCCs as nonlinear drivers, and how STC techniques can be mapped directly onto electron emission dynamics. These advances establish gas-filled waveguides as programmable interfaces between ultrafast photonics and electron science, opening opportunities in spectroscopy, imaging, and quantum applications.

**Acknowledgement**
The authors would like to acknowledge the support from the SLAC National Accelerator Laboratory, the U.S. Department of Energy (DOE), the Office of Science, Office of Basic Energy Sciences under Contract No. DE-AC02-76SF00515, No. DE-SC0022559, No. DE-FOA-0002859, the National Science Foundation under Contract Nos. 2231334, 2431903 and 2436343, and AFOSR Contract No. FA9550-23-1-0409. The authors would like to thank the useful discussions with John Travers.

**Declarations**. The authors declare no competing interests.

**Data availability**. Data underlying the results presented in this paper are not publicly available at this time but may be obtained from the authors upon reasonable request.